\documentclass[a4paper]{jpconf}
\usepackage{graphicx}
\usepackage{wasysym} 
\usepackage{hyperref}
\usepackage{color}
\begin{document}

\def\anti_particle{\tilde}
\def\clsmethod{CL$_s$ }
\def\cl{\mathrm{CL}}

\definecolor{goodcolor}{rgb}{.9 ,.01 ,.01}%
\definecolor{badcolor}{rgb}{.01 ,.2 ,.9}%
\definecolor{normalcolor}{rgb}{.0 ,.0 ,.0}%

\title{Searches for sterile neutrinos at very short baseline reactor experiments}

\author{M Danilov$^{1,2}$}
\address{$^1$ National Research Nuclear University MEPhI (Moscow Engineering Physics Institute), Kashirskoe highway 31, Moscow, 115409, Russia}
\address{$^2$ Lebedev Physical Institute of the Russian Academy of Sciences, 53 Leninskiy Prospekt, Moscow, 119991, Russia}

\ead{danilov@lebedev.ru}

\begin{abstract}

For a long time there were 3 main experimental indications in favor of the existence of sterile neutrinos: $\anti_particle\nu_e$ appearance in the $\anti_particle\nu_\mu$ beam in the LSND experiment, $\anti_particle\nu_e$ flux deficit in comparison with theoretical expectations in reactor experiments, and $\nu_e$ deficit in calibration runs with radioactive sources in the Ga solar neutrino experiments SAGE and GALEX. All three problems can be explained by the existence of sterile neutrinos with the mass square difference in the ballpark of $1~\mathrm{eV^2}$. Recently the MiniBooNE collaboration observed electron (anti)neutrino appearance in the muon (anti)neutrino beams. The significance of the effect reaches 6.0$\sigma$ level when combined with the LSND result. Even more recently the NEUTRINO-4 collaboration claimed the observation of $\anti_particle\nu_e$ oscillations to sterile neutrinos with a significance slightly higher than 3$\sigma$. If these results are confirmed, New Physics beyond the Standard Model would be required.  More than 10 experiments are devoted to searches of sterile neutrinos. Six very short baseline reactor experiments are taking data just now. We review the present results and perspectives of these experiments.
\end{abstract}

\section{Introduction}

Oscillations of the three neutrino flavors are well established. Two mass differences and three angles describing such oscillations have been measured \cite{Fogli}. Additional light active neutrinos are excluded by the measurements of the Z boson decay width \cite{PDG}. Nevertheless, existence of additional sterile neutrinos is not excluded. Moreover, several effects observed with about $3\sigma$ significance level can be explained by active-sterile neutrino oscillations. The GALEX and SAGE Gallium experiments performed calibrations with radioactive sources and reported the ratio of numbers of observed to predicted events of $0.88\pm 0.05$ \cite{SAGE}. This deficit is the so called ``Gallium anomaly'' (GA). Mueller et al. \cite{Mueller} made new estimates of the reactor $\anti_particle\nu_e$ flux which is about 6\% higher than experimental measurements at small distances. This deficit is the so called ``Reactor antineutrino anomaly'' (RAA). Both anomalies can be explained by active-sterile neutrino oscillations at Very Short Baselines (VSBL) requiring a mass-squared difference of the order of 1~eV$^2$~ \cite{Mention2011}. The LSND collaboration reported observation of $\anti_particle\nu_\mu \rightarrow \anti_particle\nu_e$ mixing with the mass-squared difference bigger than $\sim 0.1~$eV$^2$ \cite{LSND}. The initial results of the MiniBooNE tests of this signal were inconclusive and probably indicated additional effects \cite{MiniBooNE}. However, in May, 2018 the MiniBooNE collaboration presented the $4.7\sigma$ evidence for electron (anti)neutrino appearance in the muon (anti)neutrino beams~\cite{MiniBooNE2}.
The effect significance reaches 6.0$\sigma$ when the MiniBooNE and LSND results are combined. The MINIBooNE and LSND data are consistent, however the energy spectrum of the excess does not agree too well with the sterile neutrino explanation. 

\begin{figure}[th]
\centering
\includegraphics[width=0.5\linewidth]{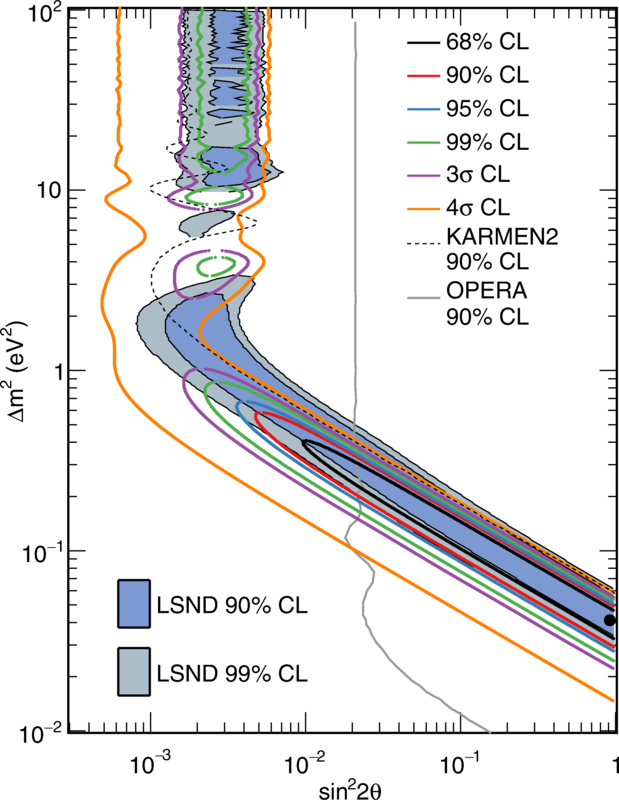}
\begin{minipage}[b]{15pc}
 \caption{\small
MiniBooNE allowed regions in neutrino mode ($12.84\times10^{20}$ POT) for events with  $200~<E^{QE}_\nu <~1250~MeV$ within a two-neutrino oscillation model. The shaded areas show the 90\% and 99\% C.L. LSND allowed regions. The black circle shows the MiniBooNE best point. Also shown are 90\% C.L. limits from the KARMEN and OPERA experiments. Figure is reproduced from~\cite{MiniBooNE2}.}
\label{MiniBooNE}
\end{minipage}
\end{figure}

 The best point in the sterile neutrino parameter space corresponds to a very large mixing ($\sin^2 2\theta=0.92$) and a small mass square difference of $\Delta m_{14}^2=0.041\mathrm{eV}^2$ (see Figure~\ref{MiniBooNE}). However, this region in the sterile neutrino parameter space is disfavored by other experiments and only a small area with larger mass square differences up to $2~{\rm eV}^2$ and smaller mixing is still allowed by the global fits~\cite{Giunti, Maltoni}.

Very recently the NEUTRINO-4 collaboration claimed the observation of the of $\anti_particle\nu_e$ oscillations to sterile neutrinos with a significance slightly larger than 3$\sigma$~\cite{NEUTRINO-4}. The measured sterile neutrino parameters are surprisingly large: $\Delta m_{14}^2=7.22\mathrm{eV}^2$ and $\sin^2(2\theta_{14})=0.35$. These values are in contradiction with the limits obtained by the reactor $\anti_particle\nu_e$ flux measurements at larger distances (see, for example \cite{DBoscillations}). However, these limits depend on the phenomenological predictions of the reactor $\anti_particle\nu_e$ flux which are model dependent.          
There are also cosmological constraints on the effective number of neutrinos~\cite{PDG,Cosmology}.
However, in several theoretical models sterile neutrinos (at least with not too large masses)are still compatible with these constraints. Details can be found in a review of sterile neutrinos \cite{WhitePaper}.

The survival probability of a reactor $\anti_particle\nu_e$ at very short distances in the 4$\nu$ mixing scenario (3 active and 1 sterile neutrino) is described by a well known expression

\begin{equation}
1-\sin^22\theta_{14}\sin^2\left(\frac {1.27\Delta m_{14}^2 [\mathrm{eV}^2] L[\mathrm m]}{E_\nu [\mathrm{MeV}]}\right).
\end{equation}

The existence of sterile neutrinos would manifest itself in distortions of the $\anti_particle \nu_e$ energy spectrum at short distances. At longer distances these distortions are smeared out and only the rate is reduced by a factor of $1-\sin^2(2\theta_{14})/2$. Measurements at only one distance from a reactor core are not sufficient since the theoretical description of the $\anti_particle \nu_e$ energy distribution is considered not to be reliable enough. The most reliable way to observe such distortions is to measure the $\anti_particle \nu_e$ spectrum with the same detector at different distances. In this case, the shape and normalization of the $\anti_particle \nu_e$ spectrum as well as the detector efficiency are canceled out. 
Usually antineutrinos are detected by means of the Inverse Beta Decay (IBD) reaction
\begin{equation}
\label{eq1}
\anti_particle{\nu}_e + p \rightarrow e^+ + n ~\mbox{with}~ E_{\anti_particle\nu} = E_{e^+} + 1.80~\mathrm{MeV}.
\end{equation}

The coherent $\anti_particle\nu_e$ scattering offers potentially another very powerful way to detect reactor $\anti_particle\nu_e$ and to search for VSBL neutrino oscillations~\cite{CONUS}.

\section{The DANSS Experiment}
The DANSS detector~\cite{DANSS} is a highly segmented plastic scintillator detector with a total volume of 1~m$^3$, surrounded with a composite shield. It consists of 2.5 thousand scintillator strips ($1 \times 4 \times 100$~cm${}^3$) with a thin ($\sim 0.2$~mm) Gd-containing surface coating. DANSS is placed under the core of the 3.1~GW$_{\rm th}$  industrial power reactor at the Kalinin Nuclear Power Plant (KNPP) 350 km NW of Moscow which provides a shielding equivalent to $\sim$~50~m of water. This shielding removes the hadronic component of the cosmic background and reduces the cosmic muon flux by a factor of 6. The very good suppression of the cosmic background and the high granularity of the detector allow DANSS to achieve a very high signal/background (S/B) ratio of more than 33 (at the 10.7~m from the reactor).

\begin{figure}[h]
\centering
\includegraphics[width=0.45\textwidth]{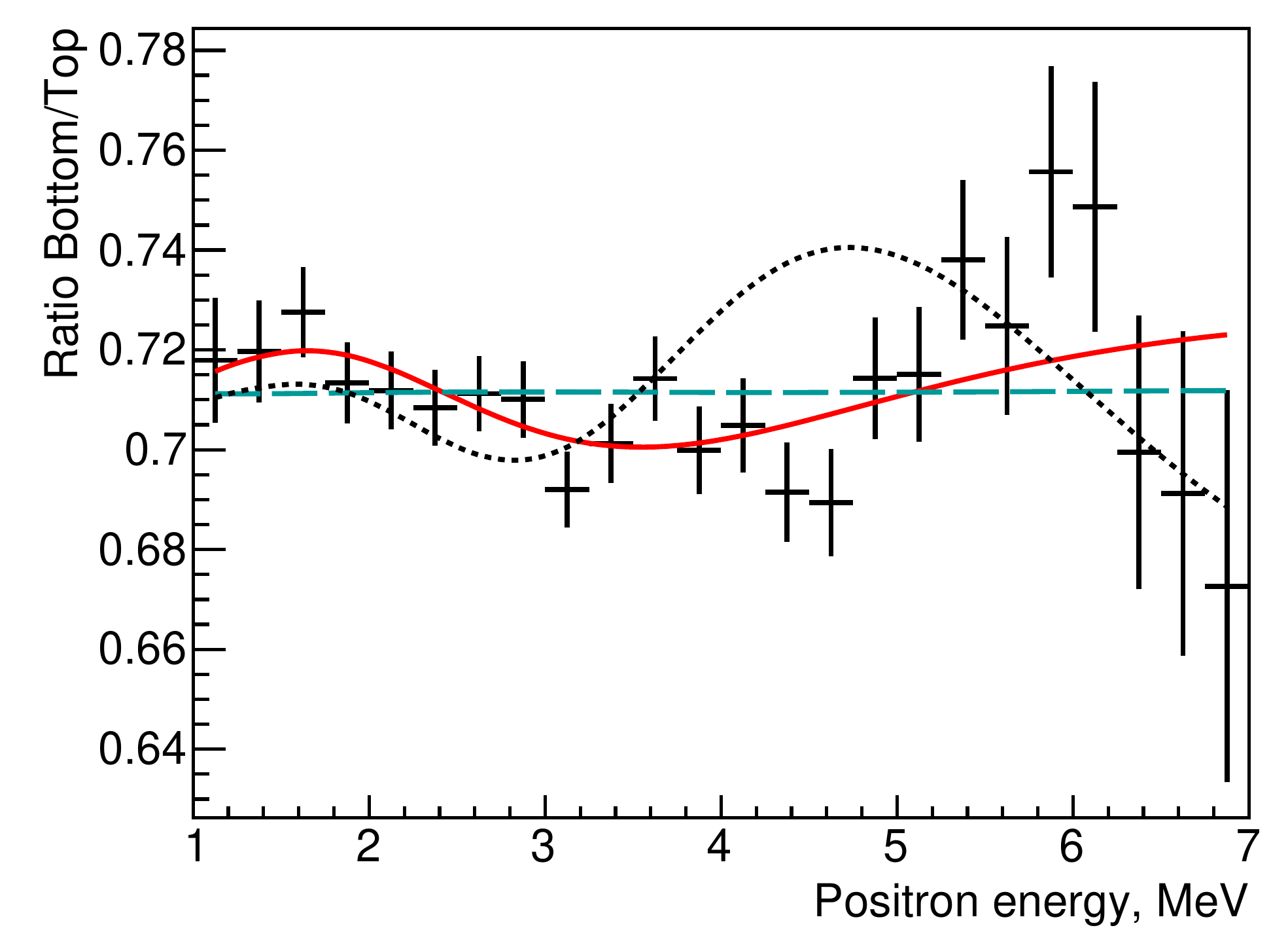}
\begin{minipage}[b]{16pc}
\caption{ \small \label{Ratio}Ratio of positron energy spectra measured by DANSS at the bottom and top detector positions (statistical errors only)~\cite{DANSSdata}. The dashed curve is the prediction for 3$\nu$ case ($\chi^2=35.0$, 24 degrees of freedom). The solid curve corresponds to the best fit in the $4\nu$ mixing scenario ($\chi^2 = 21.9$, $\sin^22\theta_{14} = 0.05$, $\Delta m_{14}^2 = 1.4~\rm{eV}^2$). The dotted curve is the expectation for the optimum point from the RAA and GA fit \cite{Mention2011} ($\chi^2=83$, $\sin^22\theta_{14}=0.14$, $\Delta m_{14}^2 = 2.3~\rm{eV}^2$)}
\end{minipage}
\end{figure}

The DANSS detector is installed on a movable platform and measures $\anti_particle \nu_e$ spectra at 3 distances from the reactor core center: 10.7~m, 11.7~m, and 12.7~m to the detector center. 
The detector positions are changed typically 3 times a week. The size of the reactor core is quite big (3.7 m in height and 3.2 m in diameter) which leads to the smearing of the oscillation pattern. This drawback is compensated by a high $\anti_particle{\nu_e}$  flux which allows DANSS to detect almost 5 thousand $\anti_particle\nu_e$ at a distance of 10.7~m. 
The energy resolution of the DANSS detector is very modest ($\sigma_E/E \sim 34\%$ at $E=1$~MeV). This leads to additional smearing of the oscillation pattern, comparable with the smearing due to the large reactor core size.
The DANSS experiment compares the positron energy spectra at the two distances from the reactor core (10.7~m and 12.7~m). The IBD counting rate is not used in the analysis (although it is consistent with the no oscillation hypothesis). This is the most conservative approach which does not depend on the predicted normalization and shape of the reactor $\anti_particle\nu_e$ spectrum as well as on the detector efficiency.
Figure~\ref{Ratio} shows the ratio of positron energy spectra at the bottom and top detector positions. 

\begin{figure}[h]
\centering
\includegraphics[width=0.5\textwidth]{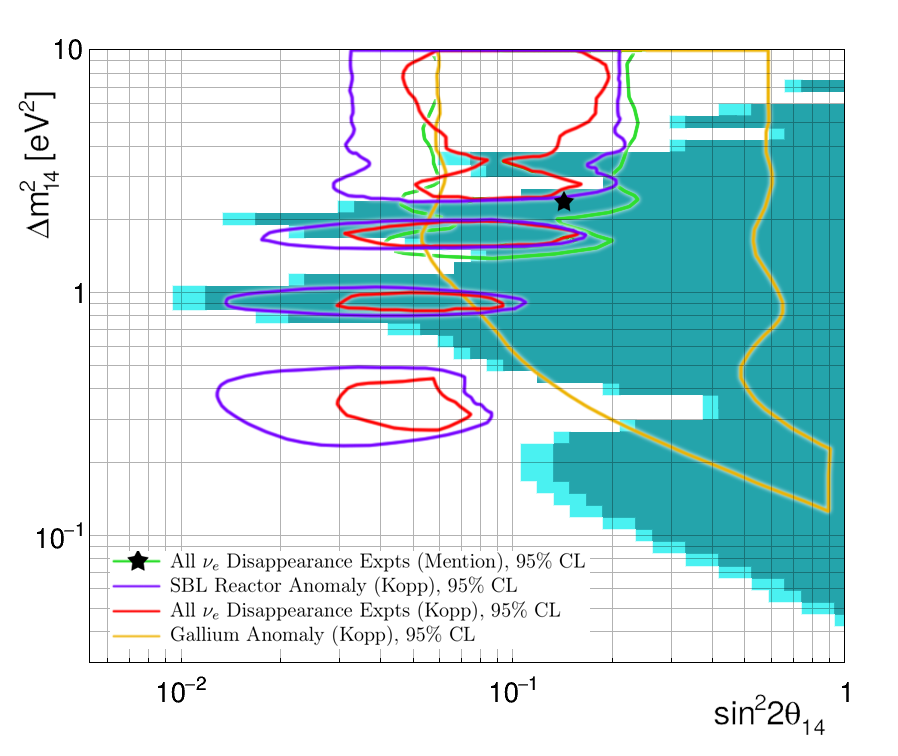}
\begin{minipage}[b]{16pc}
\caption{\small \label{Exclusion} 90\% (cyan) and 95\% (dark cyan) CL exclusion area obtained by DANSS in the $\Delta m_{14}^2,~\sin^22\theta_{14}$ parameter space~\cite{DANSSdata}. Curves show allowed regions from neutrino disappearance experiments~\cite{Mention2011, contours}, and the star is the best point from the RAA and GA fit~\cite{Mention2011}.}
\end{minipage}
\end{figure}

The optimum point of the RAA and GA fit is clearly excluded.
Figure~\ref{Exclusion} shows the 90\% and 95\% Confidence Level (CL)area excluded by DANSS in the $\Delta m_{14}^2,~\sin^22\theta_{14}$ plane. 
The excluded area covers a large fraction of regions indicated by the GA and RAA. In particular, the most preferred point $\Delta m_{14}^2=2.3~\rm{eV}^2,~\sin^22\theta_{14} =0.14$~\cite{Mention2011} is excluded at more than 5$\sigma$ CL.

The DANSS experiment continues data taking. It detected already more than 2 million IBD events. The stability of the detector is demonstrated by the perfect agreement within $\pm1.5\%$ between the reactor power and IBD rate during more than one year of operation~\cite{LaThuile}.
The DANSS experiment has the following advantages:
\begin{itemize} 
\item A very high IBD counting rate of almost 5 thousand events/day;
\item A very high S/B ratio of more than 33 at 10.7~m distance;
\item A movable 3 times a week detector;
\item A completely model independent analysis;
\item A very high detector granularity with 3D reconstruction;
\item A continuous on-line calibration with cosmic muons.

However, it has also very serious drawbacks: 
\item A very large size of the reactor core;
\item A very modest energy resolution of ($\sigma_E/E\sim 34\%$ at 1~MeV).
\end{itemize}
The DANSS experiment plans to improve considerably the energy resolution by $\sim80\%$, to increase the detector volume by $\sim40\%$, and to decrease the closest distance to the reactor core by 30~cm ($\sim15\%$ increase in the detector distance variation).

\section{The NEOS Experiment}
The NEOS detector~\cite{NEOS} has  a 1008L inner volume filled with a 0.5\% Gd-doped liquid scintillator. It is installed in the tendon gallery of 2.8~GW$_{th}$ reactor unit 5 of the Hanbit Nuclear Power Complex in Yeong-gwang, Korea at $23.7\pm 0.3$ m from the center of the reactor core. The minimum overburden with the ground and building structures corresponds to twenty meters of water equivalent. The active core size of the unit 5 is 3.1 m in diameter, 3.8 m in height. The IBD counting rate of 1976 events/day is quite high. The 73\% of background caused by the scattering and subsequent capture of fast neutrons is rejected using a pulse shape discrimination(PSD). Together with the sizable overburden this allows to achieve a very good S/B ratio of 22.

\begin{figure}[th]
\centering
\includegraphics[width=0.45\linewidth]{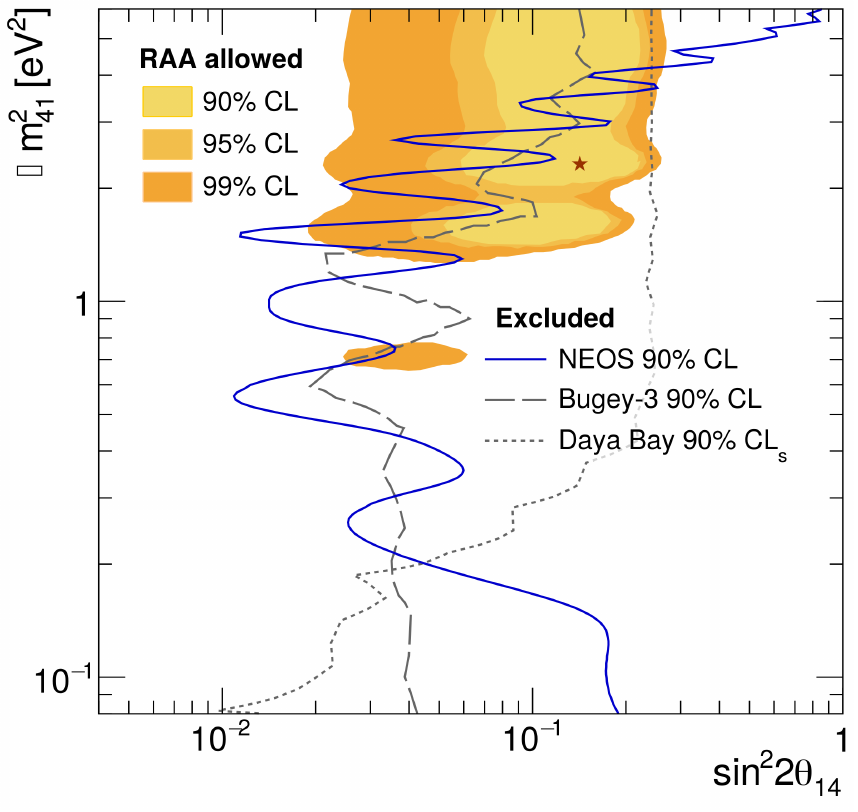}
\begin{minipage}[b]{15pc}
 \caption{\footnotesize  Exclusion curves for $3 + 1$ neutrino oscillations. The solid-blue curve is 90\%CL exclusion contours based on the comparison with the Daya Bay spectrum, the dashed-gray curve is the Bugey-3 90\% CL result~\cite{Bugey}. The dotted curve shows the Daya Bay 90\% \clsmethod result~\cite{DBoscillations}. The shaded area is the allowed region from the reactor antineutrino anomaly fit and the star is its optimum point~\cite{Mention2011}.
Figure is reproduced from~\cite{NEOS}.}
 \label{NEOSLIM}
 \end{minipage}
\end{figure}

The NEOS collaboration normalizes its data on the spectrum measured by the Day Bay collaboration at a different reactor. Such procedure is potentially vulnerable to systematic errors. There is an indication of an oscillation pattern in the ratio of the NEOS and Day Bay spectra. However, this pattern does not agree well with the expected one from neutrino oscillations. This allows the NEOS collaboration to establish strong limits on the sterile neutrino parameters shown in Figure~\ref{NEOSLIM}.

The NEOS experiment has the following advantages:
\begin{itemize} 
\item A very high IBD counting rate of almost 2 thousand events/day;
\item A good PSD;
\item A very high S/B ratio of 22;
\item A very good energy resolution of $5\%$ at 1~MeV.
\end{itemize}

However, it has also very serious drawbacks:
\begin{itemize}  
\item A very large size of the reactor core;
\item A lack of segmentation of the detector;
\item Only one and large (23.7~m) distance from the reactor core.
\end{itemize}
The NEOS experiment has resumed data taking in September 2018. They plan to contribute significantly not only to the searches of sterile neutrinos but also to the measurements of the antineutrino energy spectrum dependence on the fuel composition.

\section{The NEUTRINO-4 Experiment}

\begin{figure}[th]
\centering
\includegraphics[width=0.9\linewidth]{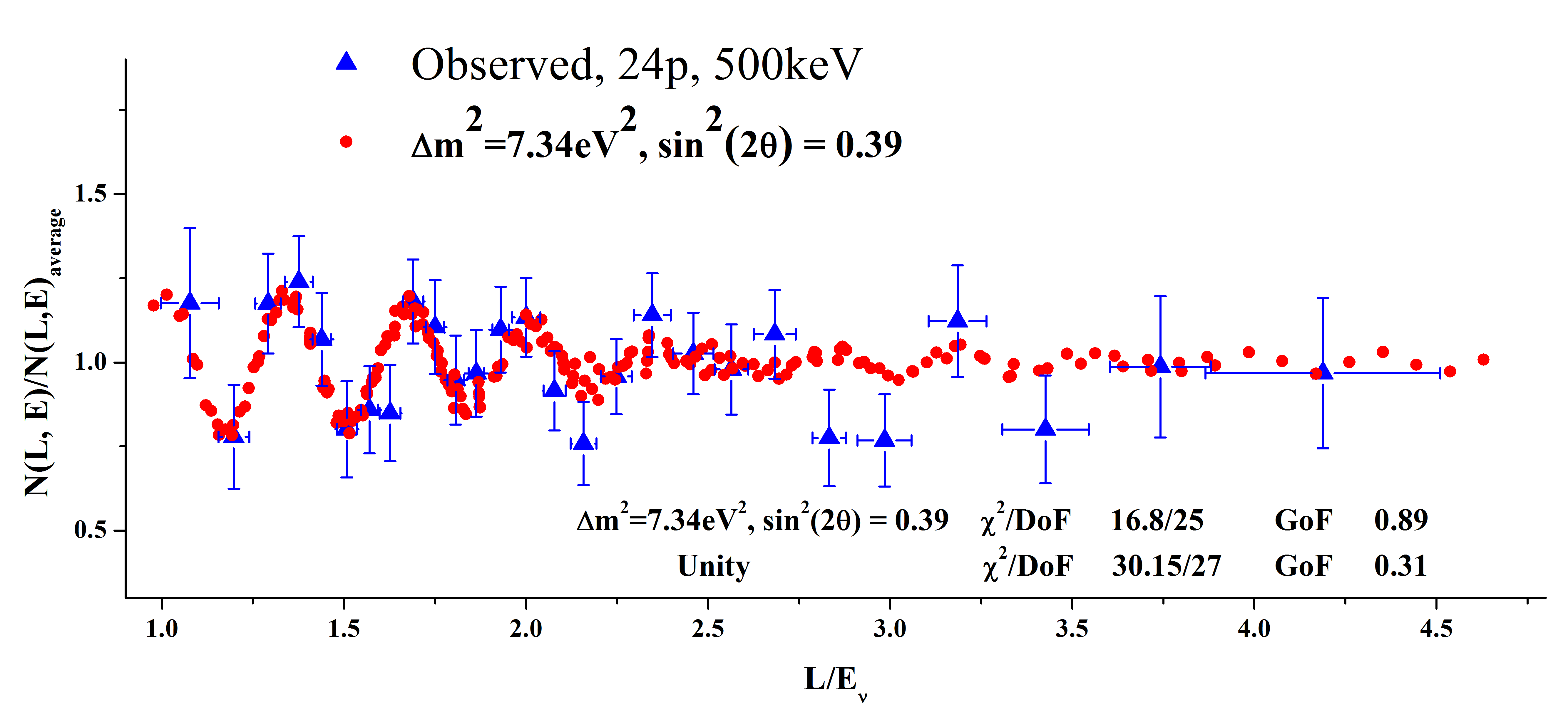}
 \caption{\footnotesize 
Normalized L/E dependence of the IBD rate in the NEUTRINO-4 experiment (triangles) and theoretical predictions (dots)~\cite{NEUTRINO-4}.}
 \label{Nu4LE}
\end{figure}

\begin{figure}[th]
\centering
\includegraphics[width=0.5\linewidth]{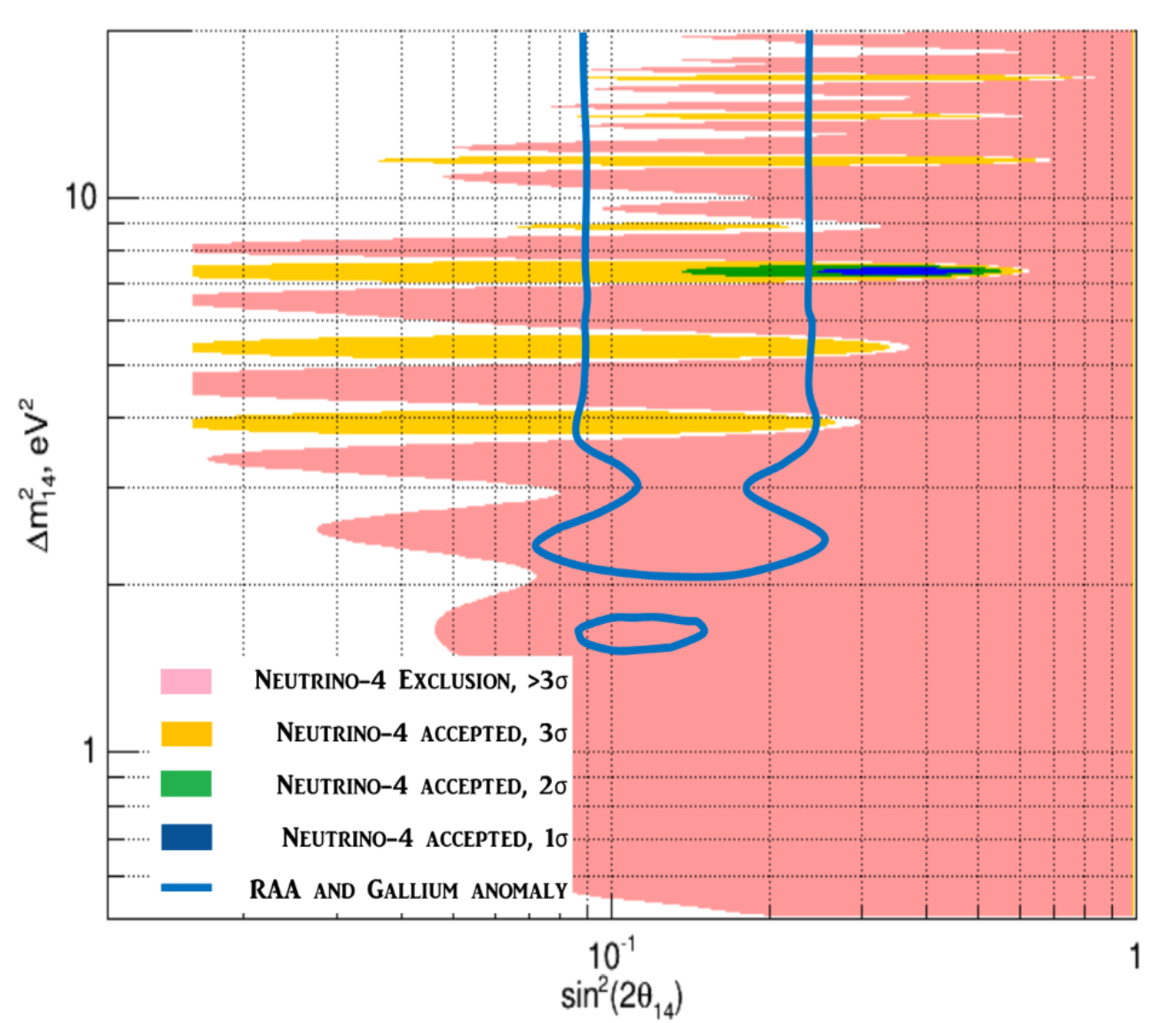}
 \caption{\footnotesize  Accepted and excluded areas for 3+1 neutrino oscillations~\cite{NEUTRINO-4}.}
 \label{Nu4LIM}
\end{figure}

The NEUTRINO-4 detector consists of 50 liquid scintillator sections with a total (fiducial) volume of $1.8(1.42)~{\rm m}^3$~\cite{NEUTRINO-4}. The detector is installed on a movable platform near a very compact ($42\times42\times35~{\rm cm}^3$) and powerful (100~MW) SM-3 research reactor at Dmitrovgrad (Russia). The distance to the reactor core is changed frequently (every 10-15 days) which allows to perform measurements in the range from 6~m to 12~m. This is an enormous asset in the control of systematic uncertainties. Unfortunately, the detector is installed almost on the ground surface with the overburden of only 3.5~m of water equivalent. There is also no PSD for the background suppression.
This leads to a very modest S/B ratio of 0.54. The energy resolution is also modest (16\% at 1~MeV).
In order to be independent from the reactor $\anti_particle\nu_e$ spectrum, NEUTRINO-4 analyses the deviations of the $\anti_particle\nu_e$ spectrum measured at different distances from the reactor core normalized on the spectrum averaged over all distances. They do not take into account the correlations between data sets but such correlations are not expected to be large. The efficiencies of different sections are assumed to be equal. The importance of this assumption is hard to estimate. It should be noted that measurements at one distance are typically averaged over several sections (with exception of the closest and the furthest distances).  This should considerably reduce the influence of possible differences in the efficiency of different sections (apart from the very important closest point).
Figure~\ref{Nu4LE} shows the obtained L/E dependence of the IBD rate normalized on the rate averaged over all distances. The data coincide nicely with the theoretical predictions which correspond to $\sin^22\theta_{14}=0.35$ and $\Delta m_{14}^2=7.22~{\rm eV}^2$. However, the theoretical predictions do not take into account the (modest) energy resolution of the detector. For example, the full width at half maximum of the 5~MeV signal should be about 840~keV which is larger even than the bin width of 500~keV used for the theoretical predictions. The example with the 5~MeV signal was selected because the significance of the oscillation effect relies mainly on the high energy part of the spectrum and the smallest distances. Incorporation of the energy resolution into the theoretical predictions should considerably smear out the oscillation pattern and to reduce the statistical significance of the effect.
Figure~\ref{Nu4LIM} shows the NEUTRINO-4 results in the $\sin^22\theta_{14},~\Delta m_{14}^2$ plane. The best point significance is larger than $3\sigma$. The obtained sterile neutrino parameters are in contradiction with the limits obtained by the reactor $\anti_particle\nu_e$ flux measurements at larger distances (see, for example~\cite{DBoscillations}). However, these limits depend on the phenomenological predictions of the reactor $\anti_particle\nu_e$ flux which are model dependent. The first results of the PROSPECT experiment~\cite{PROSPECT} are also in tension with the NEUTRINO-4 claim (see the next section) The large $\Delta m_{14}^2$ makes the disagreement with the cosmological constrains even more serious~\cite{PDG}. 

The NEUTRINO-4 experiment has the following advantages:

\begin{itemize} 
\item A very compact and powerful reactor;
\item A highly segmented and movable on-line detector;
\item A very small distance (6~m) to the reactor; 
\item A very low background from the reactor.
\end{itemize}

However, it has also many drawbacks:
\begin{itemize} 
\item A low IBD counting rate of about 200 events/day;
\item Absence of PSD;
\item A small overburden of 3.5~m;
\item A very low S/B ratio of 0.54;
\item A modest energy resolution of 16$\%$ at 1~MeV.
\end{itemize}
The NEUTRINO-4 experiment continues data taking and prepares a major modernization. The new liquid scintillator with larger amount of Gd will
decrease the neutron capture time and hence the background. The PSD will further decrease the background. The energy resolution will be considerably improved by using a two-sided readout of horizontal sections (now they are vertical with one PMT).

\section{The PROSPECT experiment}
\begin{figure}[th]
\centering
\includegraphics[width=0.5\linewidth]{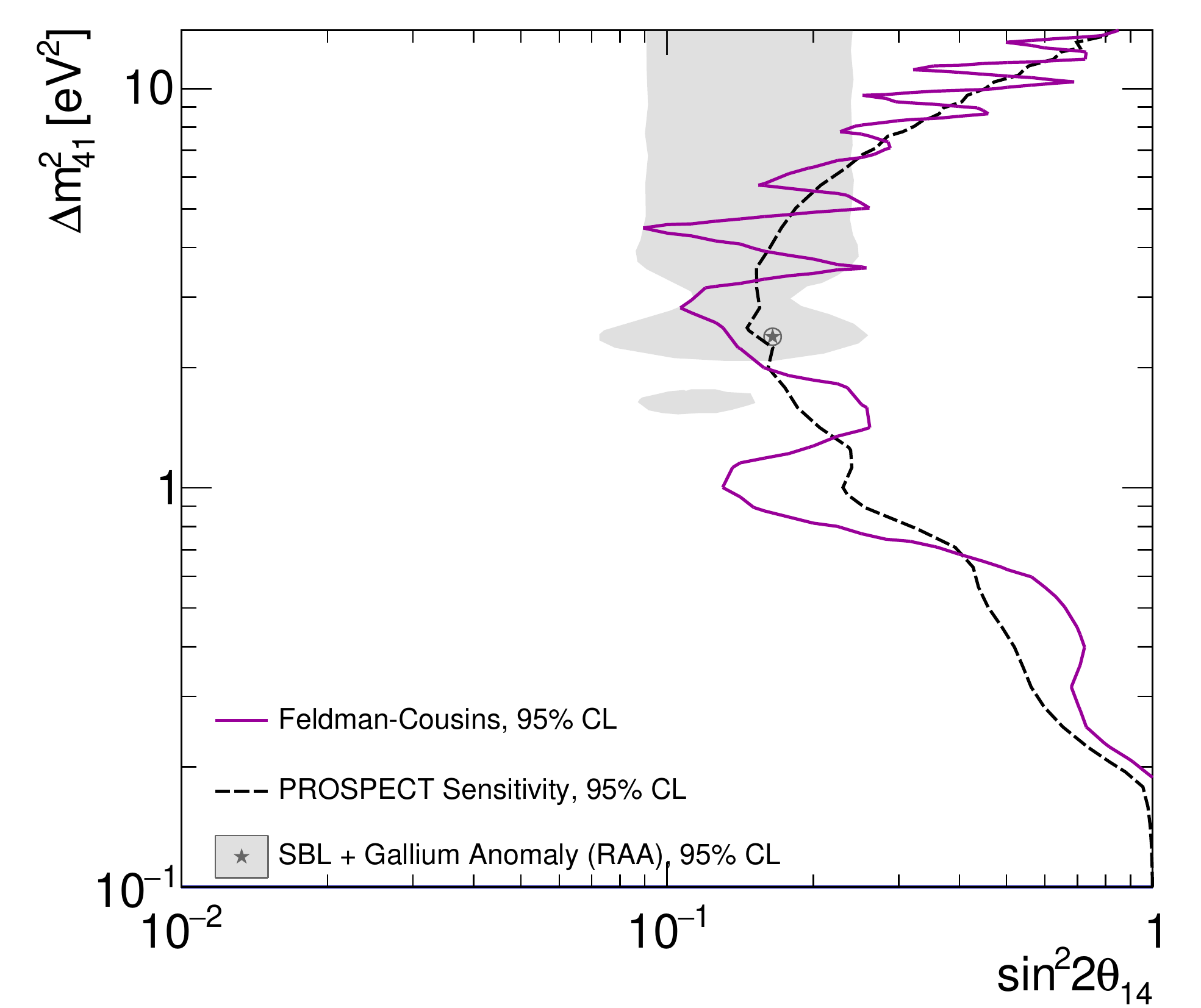}
\begin{minipage}[b]{16pc}
 \caption{\footnotesize 
Sensitivity and 95\% confidence level sterile neutrino oscillation exclusion contour from the 33 live-day PROSPECT reactor-on dataset. The best fit of the Reactor Antineutrino Anomaly~\cite{Mention2011} is disfavored at $2.2\sigma$ confidence level. 
Figure is reproduced from~\cite{PROSPECT}.} \label{PROSPECT_LIM}
\end{minipage}
\end{figure}

The PROSPECT detector is a 2.0~m$\times$1.6~m$\times$1.2~m rectangular volume containing $\sim4$ tons of pulse shape discriminating (PSD) liquid scintillator (LS) loaded with $^6$Li to a mass fraction of 0.1\%~\cite{PROSPECT}. Thin specularly reflecting panels divide the LS volume into an 11$\times$14 two-dimensional array of 154 optically isolated rectangular segments (14.5~cm$\times$14.5~cm$\times$117.6~cm) read out by two PMT each. The PROSPECT detector is installed at the High Flux Isotope Reactor (HFIR) at Oak Ridge National Laboratory practically at the earth's surface (less than one meter-water-equivalent of vertical concrete overburden). Nevertheless, a very good PSD and 3D reconstruction of events allowed PROSPECT to achieve a decent S/B ratio of 1.36. A relatively high power of the reactor (85~MW), the large detector ($\sim4$~ton), a small distance from the reactor (6.7~m), and a large detection efficiency allow PROSPECT to collect 771 IBD events per day.
In order to be independent from the $\anti_particle\nu_e$ spectrum, PROSPECT uses ratios of the measured IBD spectra at different baselines to the baseline-integrated measured spectrum. 
The measured ratios are consistent with the no oscillation hypothesis (flat behavior). This allows to exclude a sizable part of the sterile neutrino parameters (see Figure~\ref{PROSPECT_LIM}). The obtained limit on the $\sin^22\theta_{14}$ at $\Delta m_{14}^2=7.2$~eV$^2$ (preferred by the NEUTRINO-4 experiment) is smaller than the NEUTRINO-4 best point. So, there is a tension between these results but errors are still too large to draw the final conclusion.

The PROSPECT experiment has many advantages:
\begin{itemize} 
\item A very compact and powerful reactor;
\item A large (4 tons) segmented detector with 3D reconstruction;
\item A very small distance (6.7~m) to the reactor; 
\item A high IBD rate of 771 events/day;
\item A PSD of the background;
\item A very good energy resolution of 4.5~\% at 1~MeV.
\end{itemize}
There is essentially only one drawback: the small overburden which leads to a modest S/B ratio of 1.32.

\section{The SoLid experiment}

The SoLid (Phase 1) detector consists of 12800 cells which are made of a cube of polyvinyltoluene (PVT) of (5$\times$5$\times$5)~cm$^3$ in dimension, of which one or more faces are covered with thin sheets of $^6$LiF:ZnS(Ag) to capture and detect neutrons~\cite{SoLid}. The SoLid detector is installed at a distance of $\sim$~6~m of the SCK$\cdot$CEN BR2 research reactor in Belgium operated at 60~MW thermal power. The extremely high segmentation and good PSD resulted in the S/B ratio of $\sim$~3.
The SoLid experiment has the following advantages:
\begin{itemize} 
\item A very compact reactor;
\item A highly segmented detector with 3D reconstruction;
\item A very small distance (6-9~m) to the reactor; 
\item A PSD of the background;
\item A very elaborated calibration system.
\end{itemize}
The main drawback of SoLid is the modest energy resolution of 14\% at 1~MeV. There is also a big challenge of calibration of the enormous number of the detector cells.

\section{The STEREO experiment}
\begin{figure}[th]
\centering
\includegraphics[width=0.45\linewidth]{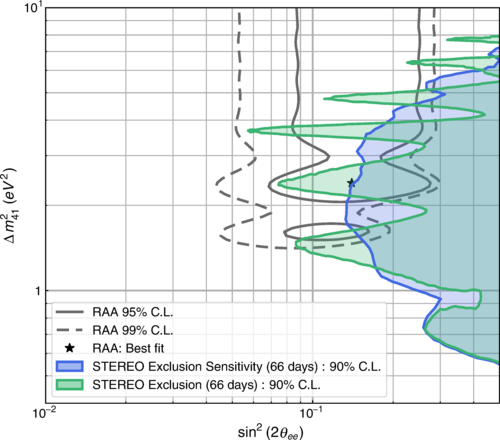}
 \caption{\footnotesize 
Exclusion contour of the oscillation parameter space. The RAA values and contours are from~\cite{Mention2011}. Figure is reproduced from~\cite{STEREO}.}
 \label{STEREO}
\end{figure}

The STEREO detector consist of six optically separated cells of the target volume filled with a gadolinium (Gd) loaded liquid scintillator for a total of almost 2~m$^3$, surrounded by passive and active shielding~\cite{STEREO}. The cells are read out from the top by four PMT per cell. STEREO is installed at the High Flux Reactor of the Institute Laue-Langevin. The cell distances from the core range from 9.4~m to 11.1~m.
Elaborated PSD technique is used to suppress the background from fast neutrons. Nevertheless, the achieved S/B ratio of 0.9 is quite modest.
STEREO uses in the analysis the ratios of prompt signal spectra of different cells to that of the first cell. This approach is independent from the $\anti_particle\nu_e$ spectrum and normalization. Already with 66 days of reactor turned on and 138 days of reactor turned off data STEREO managed to exclude a sizable fraction of the sterile neutrino parameter space~\cite{STEREO} (see Figure~\ref{STEREO}).  
 
The STEREO experiment has the following advantages:
\begin{itemize} 
\item A compact reactor core;
\item An elaborated PSD of the background.
\end{itemize}
However, there are also drawbacks:
\begin{itemize} 
\item A low S/B ratio of 0.9;
\item A modest energy resolution of 9$\%$ at 0.835~MeV.
\end{itemize}

\section{Summary}
All six VSBL experiments which are now taking data have advantages and disadvantages. They are summarized in Table~\ref{Table}.
The red color indicates advantageous parameters while problematic parameters are indicated by the blue color. The experiments DANSS and NEOS at industrial reactors benefit form a high counting rate up to 5000~$\anti_particle\nu_e$/day. They have the highest sensitivity at $\Delta m_{14}^2=$~(1-2)~eV$^2$. The sensitivity reaches regions with $\sin^22\theta_{14} < 0.01$. However, at larger $\Delta m_{14}^2$ oscillations are averaged out already inside the large reactor core and the sensitivity deteriorates. Here the experiments at the research reactors in particular PROSPECT and modernized NEUTRINO-4 have a higher sensitivity. A compilation of sensitivities of several experiments can be found in~\cite{Giunti}.
The current generation of VSBL reactor experiments will soon test the NEUTRINO-4 claim of the observation of $\anti_particle\nu_e$ oscillations to sterile neutrinos.
If confirmed this result would require New Physics beyond the Standard Model.
The comparison of the strong limits on muon (anti)neutrino disappearance with the limits on $\anti_particle\nu_e$ disappearance strongly disfavors the possibility to simultaneously explain the LSND/MiniBooNE excess in the ($\anti_particle\nu_e$)$\nu_e$ appearance and RAA/GA deficit by sterile neutrinos~\cite{Giunti,Maltoni}. However, the NEUTRINO-4 result was not included into these global fits. Forthcoming results from the running VSBL reactor experiment will soon clarify the situation.

\begin{table}[h]
\caption{\label{Table}Parameters of the VSBL experiments} 
\begin{center}
\lineup
\begin{tabular}{*{7}{c}}
\br                              
 & DANSS & NEOS & NEUTRINO-4 & PROSPECT & SoLid & STEREO\\
\mr
Power [MW] & \textcolor{goodcolor}{3100} & \textcolor{goodcolor}{2815} & \textcolor{goodcolor}{100} & \textcolor{goodcolor}{85} & 50-80 & 58\\
Core size [cm] & \textcolor{badcolor}{$\diameter = 320$}  & \textcolor{badcolor}{$\diameter = 310$}  & \textcolor{goodcolor}{$42\times42$} & \textcolor{goodcolor}{$\diameter = 51$} & \textcolor{goodcolor}{$\diameter = 50$} & \textcolor{goodcolor}{$\diameter = 40$}\\
 &\textcolor{badcolor}{$h = 370$}  & \textcolor{badcolor}{$h = 380$} & \textcolor{goodcolor}{$h = 35$} & \textcolor{goodcolor}{$h = 44$} & \textcolor{goodcolor}{$h = 90$} & \textcolor{goodcolor}{$h = 80 $}\\
Overburden [mwe] & \textcolor{goodcolor}{50} & \textcolor{goodcolor}{20} & 3.5 & $<1$   & 10  & 15  \\
Distance [m] & 10.7-12.7 & \textcolor{badcolor}{24} & \textcolor{goodcolor}{6-12} & \textcolor{goodcolor}{7-9} & \textcolor{goodcolor}{6-9} & 9-11\\
 &  \textcolor{goodcolor}{movable} &  & \textcolor{goodcolor}{movable} & & & \\
IBD events/day & \textcolor{goodcolor}{5000} & \textcolor{goodcolor}{2000} & \textcolor{badcolor}{200} & 750 & $\sim$450 & 400\\
PSD & \textcolor{badcolor}{No} & \textcolor{goodcolor}{Yes} & \textcolor{badcolor}{No} & \textcolor{goodcolor}{Yes} & \textcolor{goodcolor}{Yes} & \textcolor{goodcolor}{Yes}\\
Readout & \textcolor{goodcolor}{3D} & \textcolor{badcolor}{1D} & 2D & \textcolor{goodcolor}{3D} & \textcolor{goodcolor}{3D} & 2D\\
S/B & \textcolor{goodcolor}{33} & \textcolor{goodcolor}{23} & \textcolor{badcolor}{0.54} & 1.36 & $\sim$3 & \textcolor{badcolor}{0.9}\\
$\sigma_E/E$~[\%] at 1 MeV & \textcolor{badcolor}{34} & \textcolor{goodcolor}{5} & \textcolor{badcolor}{16} & \textcolor{goodcolor}{4.5} & \textcolor{badcolor}{14} & 8\\
\br
\end{tabular}
\end{center}
\end{table}

\section*{Acknowledgments}
The work was supported by the Russian Ministry of Science and Higher
Education contract 14.W03.31.0026.

\end{document}